# Reply to comment on "Observation of subluminal twisted light in vacuum"

Frédéric Bouchard,[1] Robert W. Boyd,[1,2,3] and Ebrahim Karimi [1,2,*]

[1] *Department of Physics, University of Ottawa, 25 Templeton St., Ottawa, Ontario, K1N 6N5 Canada.*

[2] *The Max Planck Centre for Extreme and Quantum Photonics, University of Ottawa, Ottawa, Ontario, K1N 6N5, Canada.*

[3] *Institute of Optics, University of Rochester, Rochester, New York, 14627, USA.*

*Corresponding author: ekarimi@uottawa.ca

**Recently, we showed experimentally that light carrying orbital angular momentum experiences a slight subluminality under free-space propagation [1]. We thank Saari [2] for pointing out an apparent discrepancy between our theoretical results and the well-known results for the simple case of Laguerre-Gauss modes. In this reply, we note that the resolution of this apparent discrepancy is the distinction between Laguerre-Gauss modes and Hypergeometric-Gauss modes, which were used in our experiment and in our theoretical analysis, which gives rise to different subluminal effects.**

The subluminal effect observed in vacuum is straightforwardly derived from the calculation of the group velocity from the well-known formula $v_g = |\partial_\omega \nabla \Phi|^{-1}$ [1]. Already from this formula one may notice that the group velocity depends on the transverse phase profile of the beam under study. Thus, if one is provided with the appropriate beam parameters for the phase profile, the group velocity of the propagating beam can be calculated at any given point in space $(r, \phi, z)$. We did this in [1] by directly measuring the beam radius at the focusing lens with a CCD camera (2.5 mm). Up to a good approximation (see [1] for more details), we chose the phase profile of Laguerre-Gauss modes to calculate the group velocity for our propagating beam. This family of modes is chosen because Laguerre-Gauss beams with a mode index of $\ell$ carry an orbital angular momentum value of $\ell\hbar$ along their propagation direction. Due to their symmetry in the azimuthal coordinate, an expression for the group velocity may be obtained for all value of $r$ and $z$, i.e. $v_g = v_g(r, z)$ (See Fig. 1).

With our current experimental apparatus, it is not possible to directly measure the value of the group velocity at every point in space. Instead, we measure the accumulated time delay obtained by integrating over the $z$ coordinate. However, the $r$ coordinate still remains to be dealt with. We have previously mentioned that it is the phase profile of the beam that dictates the general shape of the group velocity in space. Nevertheless, paraxial beams are generally localized to a certain transverse region in space. Here, it is the intensity of the beam that determine the value of $r$ and $\phi$ for which we consider the group velocity. The transverse intensity profile of a twisted beam of light is generally given by a ring-like (doughnut) pattern, where the radius of the ring depends upon the value $\ell$. In the case of the Laguerre-Gauss basis, the scaling law for the radius of maximum intensity is given by $r_{\max}(\ell, z) = w(z)\sqrt{|\ell|/2}$, where $w(z)$ is the well-known spot size parameter. This radius of maximum intensity taken as a function of z gives us a path (hyperbola) along which the group velocity may be integrated with respect to the z coordinate.

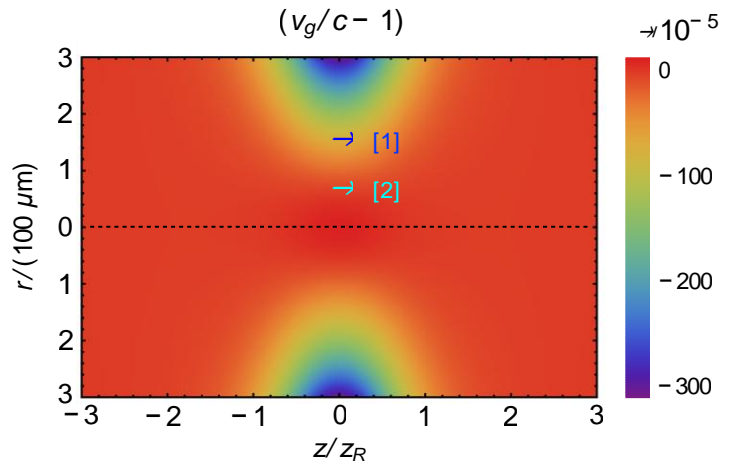

Fig. 1. Group velocity for a twisted beam of $\ell = 5$ as a function of the longitudinal and radial coordinate, i.e. $z$ and $r$. A subluminal behavior is observed near the focus and for larger radial position. The blue and cyan crosses correspond to the radial position considered in [1] and [2], respectively. A slight superluminal contribution to the group velocity is observed near the focus for smaller positions.

The apparent discrepancy between our theoretical results [1] and those of [2] is due to integration along different paths. In [2], the integration is directly performed according to the beam parameter that was previously provided, that is the beam radius at the focusing lens (2.5 mm). Results given in [2] would be perfectly reasonable if we had used Laguerre-Gauss modes in our experiment. Nevertheless, it can be seen in [1] that in order to experimentally generate the said twisted beams a pitch-fork hologram was used, without any intensity masking method. Without the use of an intensity mask [3], it is well-known that the generated beams are Hypergeometric-Gauss modes [4], not Laguerre-Gauss, which undergo a different divergence [5]. Unfortunately, intensity masking typically results in an attenuation of intensity, which we could not tolerate in our experiment because we used a nonlinear process (second-harmonic generation) in our detection scheme. Hypergeometric-Gauss modes can be expanded in the Laguerre-Gauss basis as an infinite superposition of modes with radial indices $p$ and a fixed azimuthal index $\ell$. This very point is discussed in the Supplementary Material of [1], where it is shown that it is a good approximation to calculate the group velocity using Laguerre-Gauss modes with index $\ell$. However, this is not true for the intensity profile, as it is well-known that the Hypergeometric-Gauss modes have a significantly different intensity profile. In order to reproduce the radial path of integration as well as possible in our theoretical analysis, we experimentally measured the beam radius at the focus. Finally, we used the formula given above for the maximum of intensity of the twisted modes with an effective spot size $w'(z)$, which we deduced directly from experimental results as stated in [1] (effective beam radius of 100 $\mu$m), i.e. by $r_{\max}(\ell, z) = w'(z)\sqrt{|\ell|/2}$. This point also is made in the Supplementary Material of [1] by showing a slightly different scaling law for one family of Hypergeometric-Gauss modes. Finally, it can be seen in Fig. 1 that close to the focus ($z$ and $r \to 0$), there is always a subluminal and a superluminal contribution to the group velocity, as was pointed out in [2]. However, for slightly larger radii ($r > 0.07$ mm), the group velocity is seen to go rather subluminal, with a single minimum at the focus ($z = 0$). Moreover, it can be seen from Fig. 1 that, for the radius considered in [1], it is possible to reach the same group velocity values as the one reported in [1].

In summary, we have further explained the origin of the theoretical and experimental results reported in [1] by outlining the contribution from the phase and the intensity in the calculation of subluminal group velocities. Moreover, we have clarified the nature of the discrepancy in group velocities reported in [1] and [2] by considering different families of twisted light modes, i.e. Laguerre-Gauss and Hypergeometric-Gauss modes.

**Funding.** This work is supported by the Canada Research Chairs (CRC) Program and Canada Foundation for Innovation (CFI). F.B. acknowledges the support of the Vanier Canada Graduate Scholarships Program of the Natural Sciences and Engineering Research Council of Canada (NSERC).